\documentclass[conference]{IEEEtran}
\IEEEoverridecommandlockouts
\usepackage{cite}
\usepackage{amsmath,amssymb,amsfonts}
\usepackage{algorithmic}
\usepackage{graphicx}
\usepackage{textcomp}
\usepackage{xcolor}
\def\BibTeX{{\rm B\kern-.05em{\sc i\kern-.025em b}\kern-.08em
    T\kern-.1667em\lower.7ex\hbox{E}\kern-.125emX}}

\usepackage{array}
\usepackage[caption=false,font=normalsize,labelfont=sf,textfont=sf]{subfig}
\usepackage{textcomp}
\usepackage{stfloats}
\usepackage{url}
\usepackage{verbatim}
\usepackage{graphicx}
\usepackage{siunitx}
\usepackage[version=4]{mhchem}
\usepackage{chemist}

\begin{document}


\title{Biophysical Model for Signal-Embedded Droplet Soaking into 2D Cell Culture\\
}

\author{\IEEEauthorblockN{1\textsuperscript{st} Ibrahim Isik}
\IEEEauthorblockA{\textit{School of Engineering} \\
\textit{University of Warwick}\\
Coventry, UK\\
\textit{Inonu University}\\
Malatya, Turkiye\\
ibrahim.isik@warwick.ac.uk}
\and
\IEEEauthorblockN{2\textsuperscript{nd} Hamidreza Arjmandi}
\IEEEauthorblockA{\textit{School of Engineering} \\
\textit{University of Warwick}\\
Coventry, UK \\
hamidreza.arjmandi@warwick.ac.uk}
\and
\IEEEauthorblockN{3\textsuperscript{rd} Christophe Corre}
\IEEEauthorblockA{\textit{Life Sciences} \\
\textit{University of Warwick}\\
Coventry, UK \\
c.corre@warwick.ac.uk}
\and
\IEEEauthorblockN{4\textsuperscript{th} Adam Noel}
\IEEEauthorblockA{\textit{School of Engineering} \\
\textit{University of Warwick}\\
Coventry, UK \\
adam.noel@warwick.ac.uk}

}
\maketitle
\begin{abstract}
 Using agar plates hosting a 2D cell population stimulated with signaling molecules is crucial for experiments such as gene regulation and drug discovery in a wide range of biological studies. In this paper, a biophysical model is proposed that incorporates droplet soaking, diffusion of molecules within agar, cell growth over an agar surface, and absorption of signaling molecules by cells. The proposed model describes the channel response and provides valuable insights for designing experiments more efficiently and accurately. The molecule release rate due to droplet soaking into agar, which is characterized and modeled as the source term for the diffusion model, is derived. Furthermore, cell growth is considered over the surface, which dictates the dynamics of signaling molecule reactions and leads to a variable boundary condition.  As a case study, genetically-modified $E.\,coli$ bacteria are spread over the surface of agar and Isopropyl-beta-D thiogalactopyranoside (IPTG) is considered as a signaling molecule. IPTG  droplets are dropped onto the bacteria-covered agar surface. The parameters for the IPTG molecule release rate as a diffusion source into the agar are estimated from this experiment. Then, a particle-based simulator is used to obtain the spatio-temporal profile of the signaling molecules received by the surface bacteria. The results indicate that the number of molecules reacting with or absorbed by bacteria at different locations on the surface could be widely different, which highlights the importance of taking this variation into account for biological inferences.
\end{abstract}





\section{Introduction}
Multidisciplinary approaches are being used more frequently in modern life sciences research, where traditional experimental work is closely linked with theoretical tools like biophysical modeling and computer simulation. Biophysical models, which simulate biological systems using mathematical formalizations of their physical characteristics, offer a promising approach to predict and design these experiments. An effective model should predict trials that can be tested experimentally and recommend future experiments, minimizing the requirement for difficult, expensive, or time-consuming lab work \cite{montes2019mathematical}. The $in\, vitro$ 2D cell culture is a very common experimental paradigm for biological studies. One important class of 2D cell culture experiments involves dropping diluted (by water) signaling molecules on an agar surface that hosts the cell population. The droplet gradually soaks and releases molecules into the agar, which then diffuse and interact with the cells. Agar is a jelly-like substance consisting of $polysaccharides$ obtained from the cell walls of some species of red algae and can be mixed with water and prepared similarly to gelatin before use as a growth medium. Agar is popular for cellular studies because it has a very easy preparation process and has several unique properties such as forming a gel at a relatively low temperature \cite{croze2011migration}.

Instead of describing individual cells, cell-density models in literature generally focus on the cellular density at location $x$ and time $t$ \cite{chaturvedi2005multiscale, boromand2018jamming,croze2011migration}. For example, authors of \cite{croze2011migration} study migration of $E.\,coli$ for different  cell density ranges of agar. Expanding bacterial colonies for these ranges show unique chemotactic rings. As another example, a biophysical model of dynamic balancing between excitation and inhibition for a cortical brain network is simulated in \cite{abeysuriya2018biophysical} using inhibitory synaptic plasticity in order to dynamically achieve a spatially-local balance between excitation and inhibition. A finite element model based on Fick's second law of diffusion is developed in a two-temperature agar environment to predict the radius of the inhibition zone \cite{chandrasekar2015modeling}. In this work, concentration profiles of nisin, which is used as a food preservative produced by the bacterium $Lactococcus lactis$, are calculated to establish the minimum inhibitory concentration of nisin as a function of time and location within agar, which was shown to have good agreement with experiments.

The biophysical model for signaling in a bacterial culture has been studied increasingly by  many researchers in recent years. Characterization of quorum signaling and sensing in bacterial cultures has been one focus of research, with the aim of understanding the spatio-temporal distribution of signaling molecules and its impact on gene expression \cite{ward2001mathematical,muller2013approximating, marenda2016modeling,hense2015core}.  Diffusion sensing by single cells \cite{redfield2002quorum}, clustering of bacterial aggregates inside a biofilm matrix \cite{lui2013bacteria}, collective sensing, and responsiveness to a heterogeneous environment \cite{popat2015collective} are some of the explanations for the behavior of bacterial cultures in response to signaling molecules. 

A biophysical model is proposed in \cite{trovato2014quorum} to simulate natural biofilms and predict the time-dependent concentration field of N-acyl homoserine lactone (AHL) in an agarose matrix. By monitoring AHL concentration profiles and comparing them to computed predictions in systems with various sizes and boundary conditions, the biophysical model is confirmed. The authors of \cite{trovato2014quorum} concluded from their study that diffusion sensing contributes to cell–cell communication in biofilms and if an equation for AHL diffusion included the signal's decay, then the measured concentrations and predictions are in good agreement. The authors of \cite{ward2001mathematical,muller2013approximating} show that the diffusion of an AHL signal over macroscopic distances can trigger gene expression, creating spatial and temporal patterns that are very different from those produced by simple diffusion. In \cite{dilanji2012quorum}, the authors embedded a quorum sensing (QS) strain in a short agar lane and added an exogenous autoinducer at one end of the lane to quantify the expression of a QS reporter as a function of time and space as the autoinducer diffused along the lane. A droplet-based bacterial communication system which demonstrates the potential of the system
for genetically-programmed pattern formation and distributed computing is developed using $E.\, coli$ bacteria enclosed within sizable populations of water-in-oil emulsion droplets \cite{weitz2014communication}. 

Concentration gradients for oxygen, pH, and soluble substances including nutrients and effector chemicals as well as cellular metabolites exist within a cell culture \cite{langhans2018three}. These natural gradients are impacted by the cellular metabolism and the extracellular matrix's composition. The metabolism controls how much oxygen and other nutrients are consumed and how much waste is produced. The extracellular matrix is crucial for cell growth, movement, and other functions, and it helps cells to attach to and communicate with other nearby cells.

Cell motility, cell migration, and especially cell signaling are all impacted by molecular concentration gradients \cite{langhans2018three}. We are missing a comprehensive biophysical model for signaling molecule propagation within a 2D cell culture. The droplet soaking process and subsequent release of molecules into agar, cell growth, and diffusion have not been modeled in previous works. In this study, we formulate the diffusion model for a system that includes these processes. We propose a mathematical model for droplet soaking into agar and derive the corresponding molecule release rate into the agar, which becomes the source term for the diffusion model. Interestingly, the boundary condition (BC) at the agar surface is time-variable as the cells are growing, which leads to time-varying reaction rates. As a case study, genetically-modified $E.\, coli$ bacteria are manually spread over the surface of an agar plate, which enables them to grow. A signaling molecule Isopropyl-beta-D-thiogalactopyranoside (IPTG) is diluted by water and dropped on a specific point over the bacteria-covered agar surface. We used the IPTG molecule as a signaling molecule because it leads to fluorescence that we can measure because the bacteria which we assume are genetically-engineered to react in this way in the presence of a suitable IPTG concentration. We derive the parameters for the IPTG molecule release rate into agar based on the results from this experiment. Furthermore, we use a particle-based simulator (PBS) to analyse the spatio-temporal profile of the molecules within the agar and the number of molecules consumed by bacteria over the surface. The aim of the present study is to provide valuable insights for designing experiments more efficiently and accurately. We show that the number of molecules absorbed by cells at different locations on the surface could be widely different, which highlights the importance of taking this variation into account for biological inferences.

The rest of the paper is organized as follows. Section 2 describes the system model. Section 3 describes the proposed biophysical model for droplets soaking into agar and the diffusion of molecules within the agar. In Section 4, simulation results are given using PBS. Finally, Section 5 concludes the paper.

\section{System Model}
\begin{figure}
\centering
\includegraphics[width=3.5in]{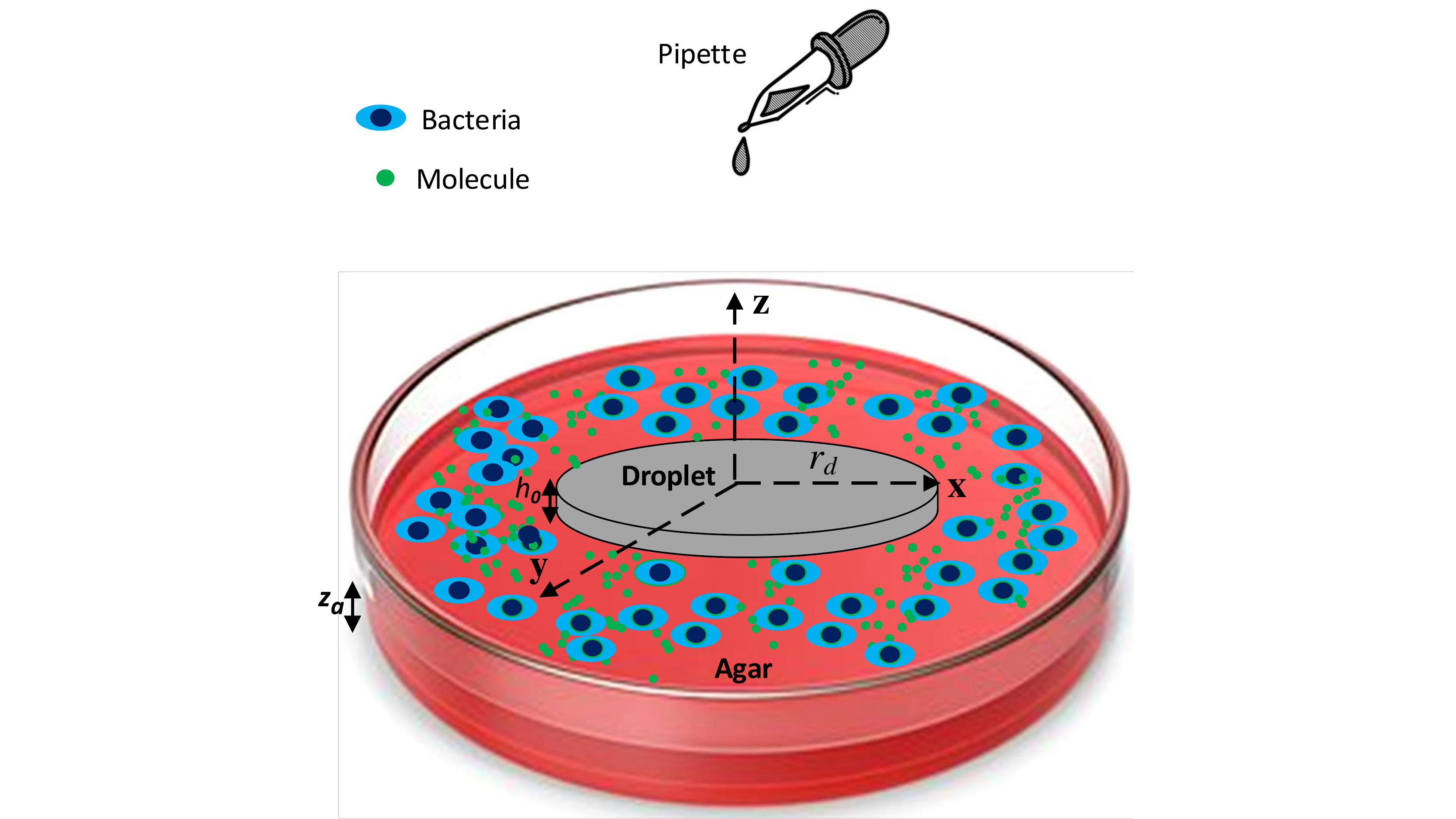}
\caption{Schematic of the system model. A droplet whose height and radius are $h_0$ and $r_d$ respectively is dropped onto the center of the agar surface whose height is $z_a$.}
\label{model}
\end{figure}

In this section, we model molecule propagation from a droplet added to a cell population over agar. We consider a cylindrical plate of radius $r_p$ and height $h_p$ filled up to height $z_a$ with agar. A cylindrical coordinate system is used whose origin is located at the center of the circular surface at the top of plate. The variables $(r,\theta, z)$ denote the radial, azimuthal, and axial coordinates. We assume that the bacteria source is diluted with water to a bacterial concentration of $C_b$ that can be measured using optical spectroscopy. From this source, $V_b$ \si{m^3} is added and spread uniformly over the whole surface of the agar.  Fig. \ref{model} depicts the schematic of the system model. We assume that the signaling molecules are also diluted with water. A droplet volume $V_d$ \si{m^3} of this solution with a concentration of $C_m$ \si{moles/m^3} of signaling molecules is considered. The droplet is dropped over the center of the agar surface and it forms a thin layer of area $A_d$ \si{m^2}. The diluted solution including the signaling molecules soaks into the agar, which has a diffusable solid/semi-solid medium, and shrinks over time, thus we write $V_d$ as $V_d(t)$. The signaling molecules diffuse through the plate and may react with cells over the surface through their diffusion. We assume that the agar's porosity enables the droplet to soak without any change in the agar volume properties.

\section{Proposed Biophysical Model}
In this section, we present the biophysical model for a droplet soaking into bacteria-covered agar and the resulting diffusion and consumption processes. First, we derive the signaling molecule release rate at the agar surface. Then, we present the differential equations that describe how the molecules diffuse through the agar and get consumed by the growing surface bacteria population.

\subsection{Droplet soaking and molecule release} 
Due to cohesion forces, liquids at rest tend to occupy the smallest possible surface area in order to achieve the highest possible energy balance \cite{berry1971molecular}. Thanks to surface tension, water spilled on the ground behaves like an elastic layer and creates a layer a few millimeters above ground level instead of occupying a large surface area.

We assume that the droplet of volume $V_{d}(t)$ \si{m^3} is dropped over the agar surface at time $t = 0$ and occupies an area of $A_d(t)$ \si{m^2} that can be measured experimentally. Therefore, the initial height of the droplet is obtained as
\begin{equation}
 h_{0}=\frac{V_{d}(0)} {A_{d}(0)}.
\label{h}
\end{equation}

We assume that the shape of the droplet is ideally a cylinder of radius $r_{d}(0)$, i.e., $A_{d}(0)=\pi r_{d}^{2}(0)$.

After it lands, the droplet begins to soak into the agar over all of the contact surface between the droplet and agar at a soaking rate of $K$ \si{m/s}. Therefore, the volume starts to decrease at a rate of $A_d(0)K$ or $\pi r_{d}^{2}(0)K$ \si{m^3/s} for an ideal cylindrical shape. By soaking over time, the droplet volume reduces. As the volume decreases, the cohesion force implies that the droplet occupies a smaller contact area with the agar surface. Thus, we assume that the droplet height $h_{0}$ remains $fixed$ while the droplet shrinks. As a result, given the droplet area of $A_{d}(t)$ at time $t$, and the soaking rate of $K$ \si{m/s}, the volume of the droplet and the volume change rate are respectively obtained as 
\begin{equation}
  V_{d}(t)=A_{d}(t)h_{0},
  \label{Vdt}
\end{equation}
and
\begin{equation}
  \frac{dV_{d}(t)}{dt}=-A_{d}(t)K.
  \label{Vdt2}
\end{equation}

Considering  (\ref{h}), (\ref{Vdt}) and (\ref{Vdt2}), it is straightforward to obtain
\begin{equation}
  A_{d}(t)=A_{d}(0)e^ {\frac{-K}{h_{0}}t}.
  \label{Adt}
\end{equation}

Given the ideal circular shape, the dynamic radius   $r_{d}(t)$ is also obtained as
\begin{equation}
  r_{d}(t)=r_{d}(0)e^ {\frac{-K}{2h_{0}}t}.
  \label{rdt}
\end{equation}

As we will show, the constant $K$ depends on the concentration of the signaling molecules in the droplet, so we denote it as $K(C_{m})$ in the rest of the paper. 

Given the concentration $C_m$ of the signaling molecules in the droplet, we have $N(t)=C_mV_{d}(t)$ molecules at time $t$ within the droplet. Considering (\ref{Vdt2}), the release rate of the molecules over the droplet area $A_{d}(t)$ is given by
\begin{equation}
  \frac{dN(t)}{dt}=C_mK(C_{m})A_{d}(0)e^ {\frac{-K}{h_{0}}t}.  
  \label{dNt}
\end{equation}

In other words, during the short interval $\Delta t$, the number 
{$\frac{dN(t)}{dt}\Delta t$} molecules are uniformly released over the area $A_{d}(t)$ occupied by the droplet.

\subsection{Diffusion model}
The molecule concentration $c(\bar{r},t)$ due to molecule release by the droplet and random molecule diffusion is described by Fick's second law of diffusion  and a source term as given by 
\begin{equation}
\frac{\partial c(\bar{r},t)}{\partial t}=D\nabla 
^{2}c(\bar{r},t)+S(\bar r ,t),
\label{diff}
\end{equation}
where $D$ is the coefficient diffusion of molecules within agar and the molecule release rate $S(\bar r,t )$ models the droplet surface molecule source where the radius of the droplet goes zero and can be obtained using the unit step function as given below,
\begin{equation}
 \begin{split}
 S(\bar{r},t)=K(C_{m})[u(t)][f_A(\bar r,t)][\delta(z)],
 \label{sourceeqn}
  \end{split}
 \end{equation}
where $f_A(\bar r,t)=1$ for all $\bar r$ in the area occupied by the droplet at time $t$, otherwise $f_A(\bar r,t)=0$, i.e., $f_A$ is an indicator function that indicates whether $\bar r$ is inside the droplet area at time $t$. The side walls and bottom of the agar plate imply the following reflective boundary conditions (BCs): 

\begin{equation}
\frac{\partial C_{m}(\bar{r},t)}{\partial r}\Bigg| _{r=r_p}=0,
\label{bound1}
\end{equation}
where $r_p$ is the plate radius, and
\begin{equation}
\frac{\partial c(\bar{r},t)}{\partial z}\Bigg| _{z=h_a}=0.
\label{bound2}
\end{equation}

 Another boundary condition is considered at the surface of the agar where the bacteria are growing and consuming signaling molecules. To model it, we need to characterize the uniform cell growth over the agar and its molecule consumption rate. For simplicity, we consider an irreversible reaction between the molecules ($\mathcal{M}$) and cells ($\mathcal{C}$) and suppose that a reaction mechanism of the following type can degrade the $\mathcal{M}$ molecules across the environment:
\begin{equation}
  \ce{\mathcal{M} + \mathcal{C} 
  \reactrarrow{0pt}{1.5cm}{\ChemForm{K_B(t)}}{}
  \mathcal{MC}},
\end{equation}
 where the $K_B(t)$ is the consumption rate of cells over the surface. This reaction mechanism assumes that there is no saturation constraint, so we can approximate it as a first-order surface reaction. Thus, the corresponding BC is defined as
\begin{equation}
\frac{\partial c(\bar{r},t)}{\partial r}\Bigg| _{z=0}=-K_B(t)(C_{m}).
\label{bound3}
\end{equation}

 Since the cells are growing over the surface, we assume that the consumption rate increases over time. We assume that the consumption rate scales linearly with the cell population, i.e., $K_B(t)=K_B(0)P_B(t)/P_B(0)$, where $P_B(t)$ denotes the cell population at time $t$. For example, within the growth stage, a population of $E.\, coli$ cells doubles every 20 min \cite{hense2015core}. Thus, given the initial consumption rate $K_B(0)$, we can estimate the time-varying consumption rate as
\begin{equation}
K_B(t)=K_B(0)2^{t/20}\, \textrm{for}\,\, t=(20, 40...)\, \textrm{min}
\label{kbt}
\end{equation}

\begin{table}[htbp]
  \caption{Parameters for the proposed biophysical model.\label{parameters}}
   \begin{tabular}{l l}
   \hline
    Parameter&Value\\
    \hline
    IPTG diffusion coefficient, $D$  & $5 \times 10^{-11} \si{m^2/s}$\\
    Droplet (transmitter) radius \\
    at start of the experiment, $r_d$ & 0.01 m\\
Plate radius, $r_p$ & 0.1 m\\

Droplet IPTG concentration, $C_m$ & (0.001, 0.01, 0.1) M or \si{moles/m^3}\\

Soaking rate, $K(C_m)$ & $(1.2\times10^7, 6.1\times10^8, 5.5\times10^9)$ \\
&\si{m/s}\\

Total number of molecules \\released into agar &  $(1.2, 3.2, 5.2)\times10^6$\\

Droplet height, $h_0$ & $3.18\times10^{-5}$ m\\

 Agar height, $z_a$ & $6\times10^{-3}$ m\\
 \hline
\end{tabular}
\end{table}

\section{Simulation Results}

\begin{figure}
\centering
\includegraphics[width=3.5in]{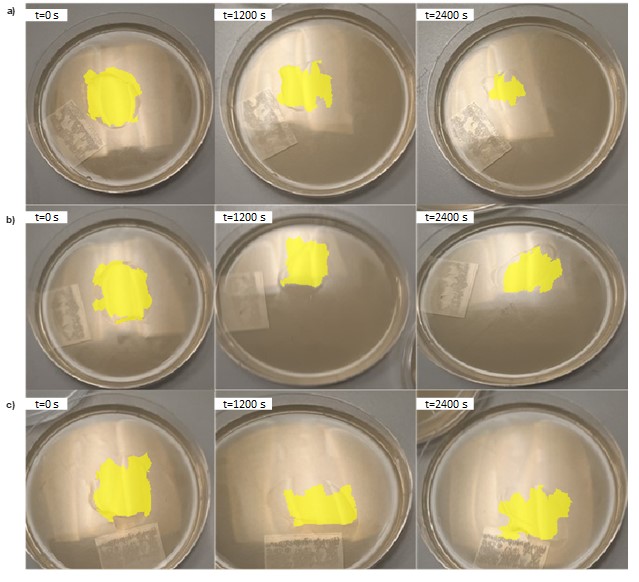}
\caption{Shrinking of the droplet (yellow) with different IPTG concentrations a) $C_m=0.1$ M, b) $C_m=0.01$ M, c) $C_m=0.001$ M dropped over surface of agar with $100\,\mu\textrm{L}$ of uniformly-spread $E.\,coli$ bacteria solution. Image highlighting with MATLAB is applied here to clearly show shrinkage of the droplet over agar.}
\label{sakig}
\end{figure}

\begin{figure}
\centering
\includegraphics[width=3.5in]{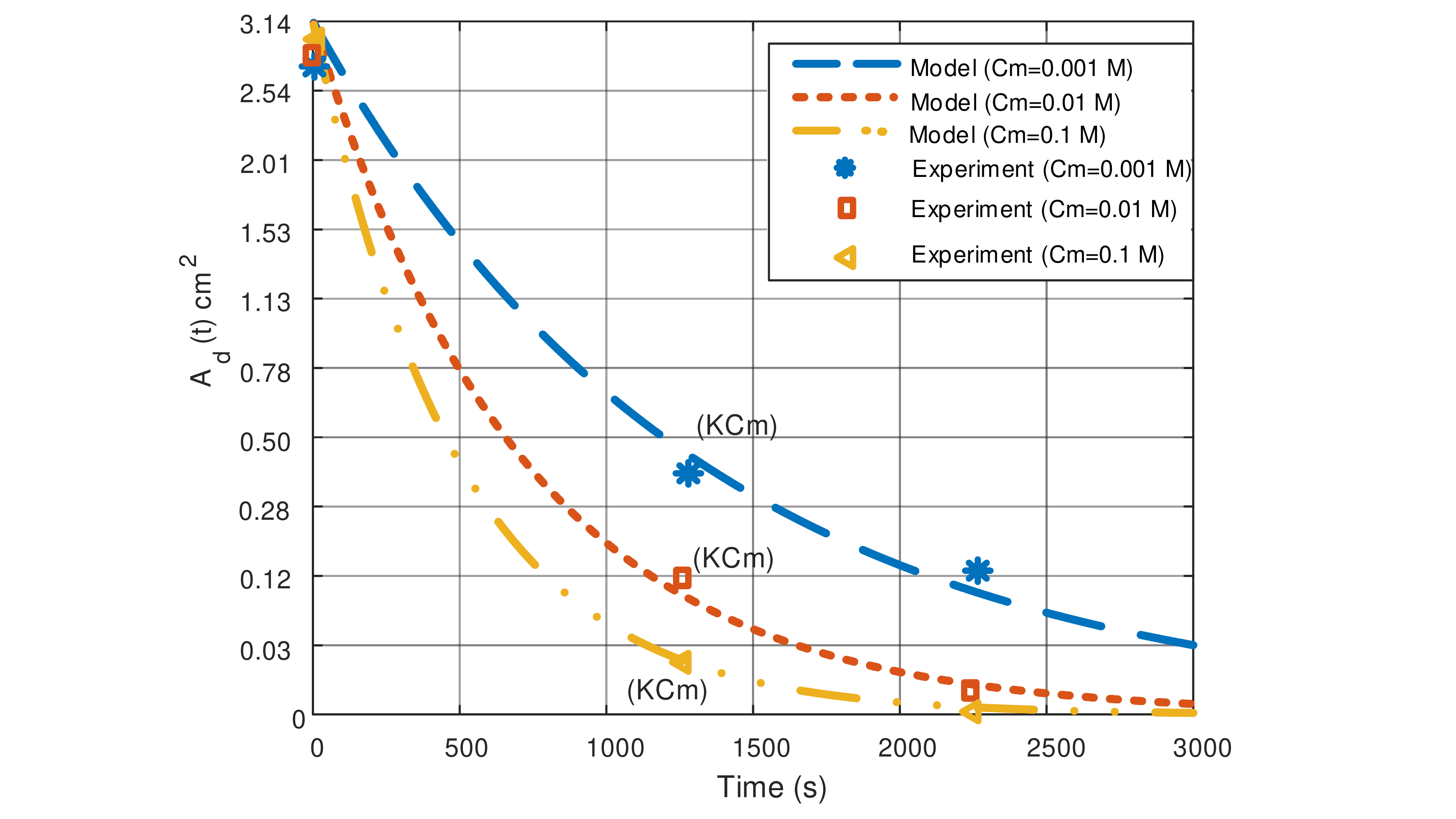}
\caption{$A_{d}(t)$ versus time for different droplet IPTG concentrations, $C_m=(0.1.0.01,0.001)$ M that soak into the agar with different rates denoted by $K(C_m)$ calculated from (\ref{rdt}) and measured from Fig. \ref{sakig}.}
\label{k}
\end{figure}
  
In this section, we first use the experimental data to extract the parameters of the model proposed for droplet soaking and characterize the release source of molecules into the agar. Then, we use a PBS to simulate the propagation of signaling molecules throughout the agar plate and their consumption by the cells growing on the surface. The shrinking occurs faster for a higher molecule concentration as shown in our experimental results of droplet soaking into agar in Fig. \ref{sakig}. Using Fig. \ref{sakig}, we can estimate the droplet soaking rate $K(C_m)$ as a function of the droplet IPTG concentration. Given the droplet volume of 10 $\mu\textrm{L}$, $h_0$ is obtained using (\ref{Vdt}). The area of the droplet was measured at times 0, 1200, and 2400 seconds for each experiment and the corresponding soaking rates $K(C_m)$ have been obtained $1.2\times10^7$ \si{m/s}, $6.1\times10^8$ \si{m/s}, and $5.5\times10^9$ \si{m/s} for IPTG concentrations of 0.1, 0.01 and 0.001 respectively using (\ref{rdt}). Initial and final values of droplet radius ($r_{d}$) are used in (\ref{rdt}) to obtain these soaking rates. Using the estimated $K(C_m)$, the droplet area is plotted in Fig \ref{k} as a function of time for each concentration $C_m$. We plot $A_{d}$ versus time to show how experiment and model results are fitted. For example, calculated soaking rates using (\ref{rdt}) for $t$=1200 s shown in Fig. \ref{k}. We estimate the droplet radius using (\ref{rdt}) by assuming that the droplet is an ideal cylinder, even though it is clear from Fig. \ref{sakig} that this is not strictly true.     

\subsection{Experimental results}

In this study, the droplet soaking rates are obtained using experimental data of Fig. \ref{sakig}. The experimental data are used to extract the parameters of the model proposed for droplet soaking and characterize the release source of molecules into the agar. During the experiment, 100 $\mu$L of concentration $E.\,coli$ liquid culture is diluted in 100 $m$L agar liquid and is kept for 24 hours to grow. 100 $\mu$L of the diluted bacteria source is spread over the surface of three agar plate with diameter $10$cm. We prepared three different concentrations  $C_m=(0.001, 0.01, 0.1)$ M of IPTG as a signaling molecule by diluting it in water. From each source, we have dropped 10 $\mu$L of the diluted IPTG over the center of the agar plates. Fig. \ref{sakig} shows the images taken from the agar surface at times $t=0, 1200, 2400$ s to measure the droplet shrinking over the surface for these three different concentrations. The coverage area of the droplet has been highlighted through an image processing code in MATLAB.

\subsection{Particle based simulation (PBS)}

The soaking rates found using the experimental results characterize the release source term for the diffusion model. A PBS  is used to simulate the proposed biophysical model for signaling molecules over the agar plate and their consumption by the bacteria. In the PBS, during the simulation, time is divided into time steps. The parameters for the biophysical
model are given in Table \ref{parameters}. To model the source term, during the short interval of $\Delta t=1000$ s, a number of molecules $C_mK(C_{m})A_{d}(0)e^ {\frac{-K}{h_{0}}t}\Delta t$ is uniformly released over the area $A_{d}(t)$ occupied by the droplet considering (\ref{dNt}). The locations of the molecules are updated in accordance with Brownian motion (.e., the displacements are Gaussian-distributed with mean 0 and variance $\sqrt{(2D \Delta t)}$ along each cartesian axis direction) at each time step $\Delta t$. We have also considered the bacteria growth over the agar surface and obtained the dynamics of the absorption rate of the bacteria over the surface. Each molecule hitting the side wall and bottom of the plate is reflected back into the environment according to the BCs (\ref{bound1}) and (\ref{bound2}). We used the PBS model to obtain the spatio-temporal profile of signaling molecules consumed by the bacteria over the surface. In other words, the number consumed is obtained using a probabilistic rate \cite{seong2020control} and the position of molecules consumed by the bacteria are saved at each time step. 

\begin{figure}
\centering

\includegraphics[width=3.3in]{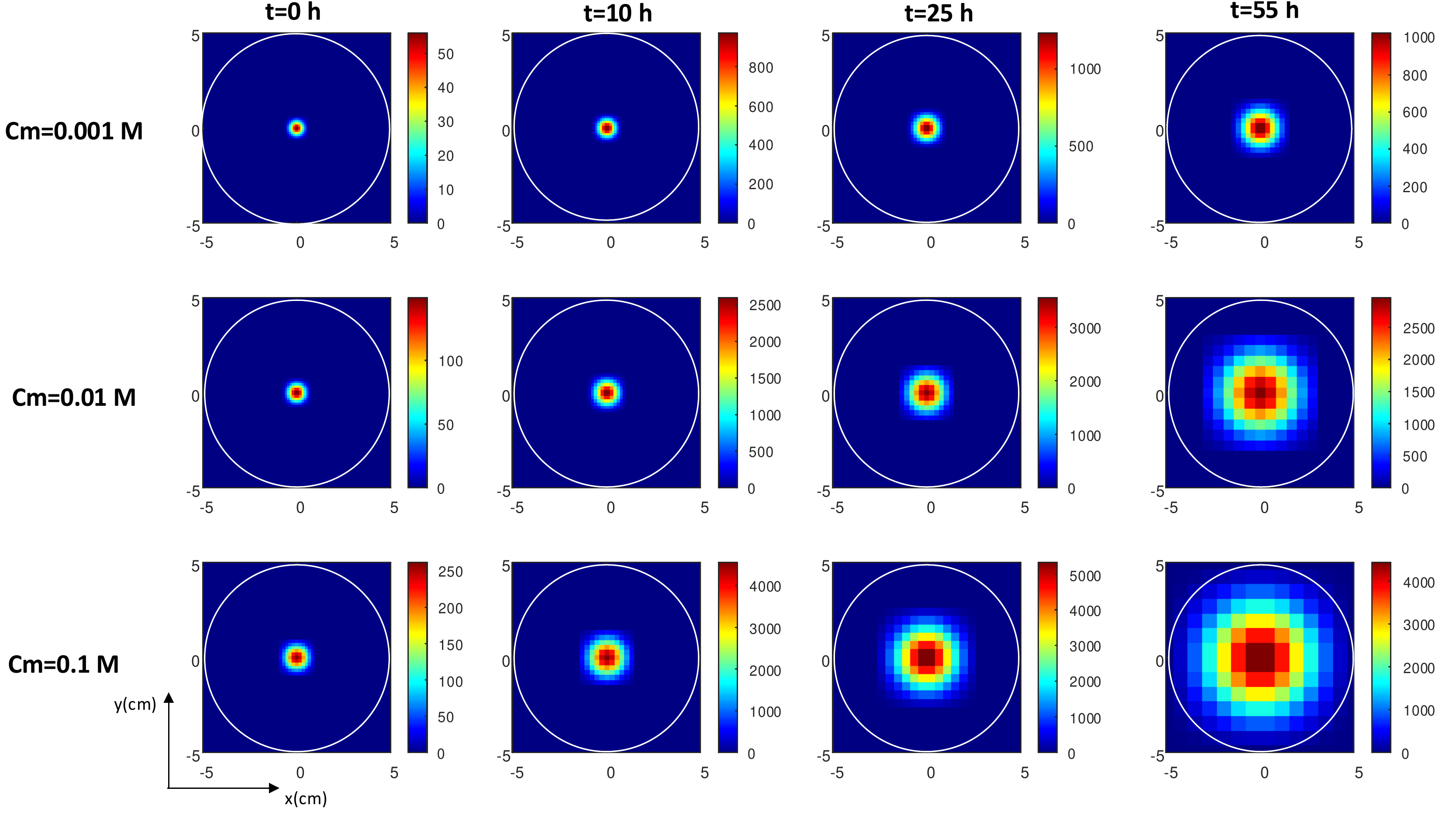}
\caption{IPTG concentration profile in 2D (x,y) (integrating over z-axis) for 3 different IPTG droplet concentrations, $C_m=(0.001, 0.01, 0.1)$ M in top, middle, and bottom rows, respectively. Each row shows 4 time samples of the concentration profile at \{0, 10, 25, 55\} hours from left to right. White circles on each picture represent the plate edge.} 
\label{mol_pos}

\end{figure}
\begin{figure}
\centering

\includegraphics[width=3.3in]{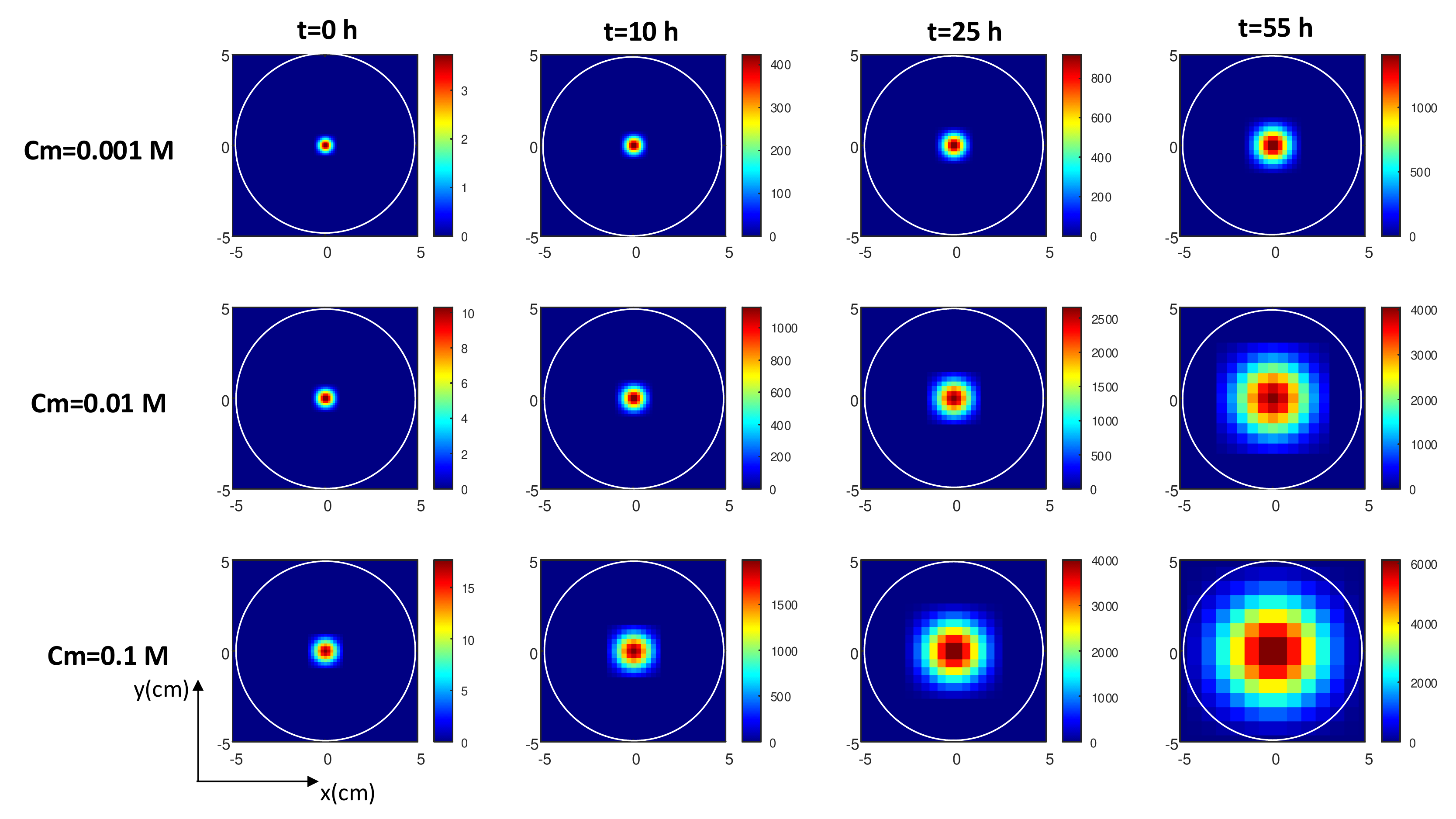}
\caption{The profile of IPTG consumed by the bacteria population over the surface for 3 different IPTG droplet concentrations, $C_m=(0.001, 0.01, 0.1)$ M in top, middle, and bottom rows, respectively. Each row shows 4 time samples of consumed amount of molecules at \{0, 10, 25, 55\} hours from left to right. White circles on each picture represent plate edge.}
\label{bac_pos}

\end{figure}
\begin{figure}
\centering
\includegraphics[width=3.5in]{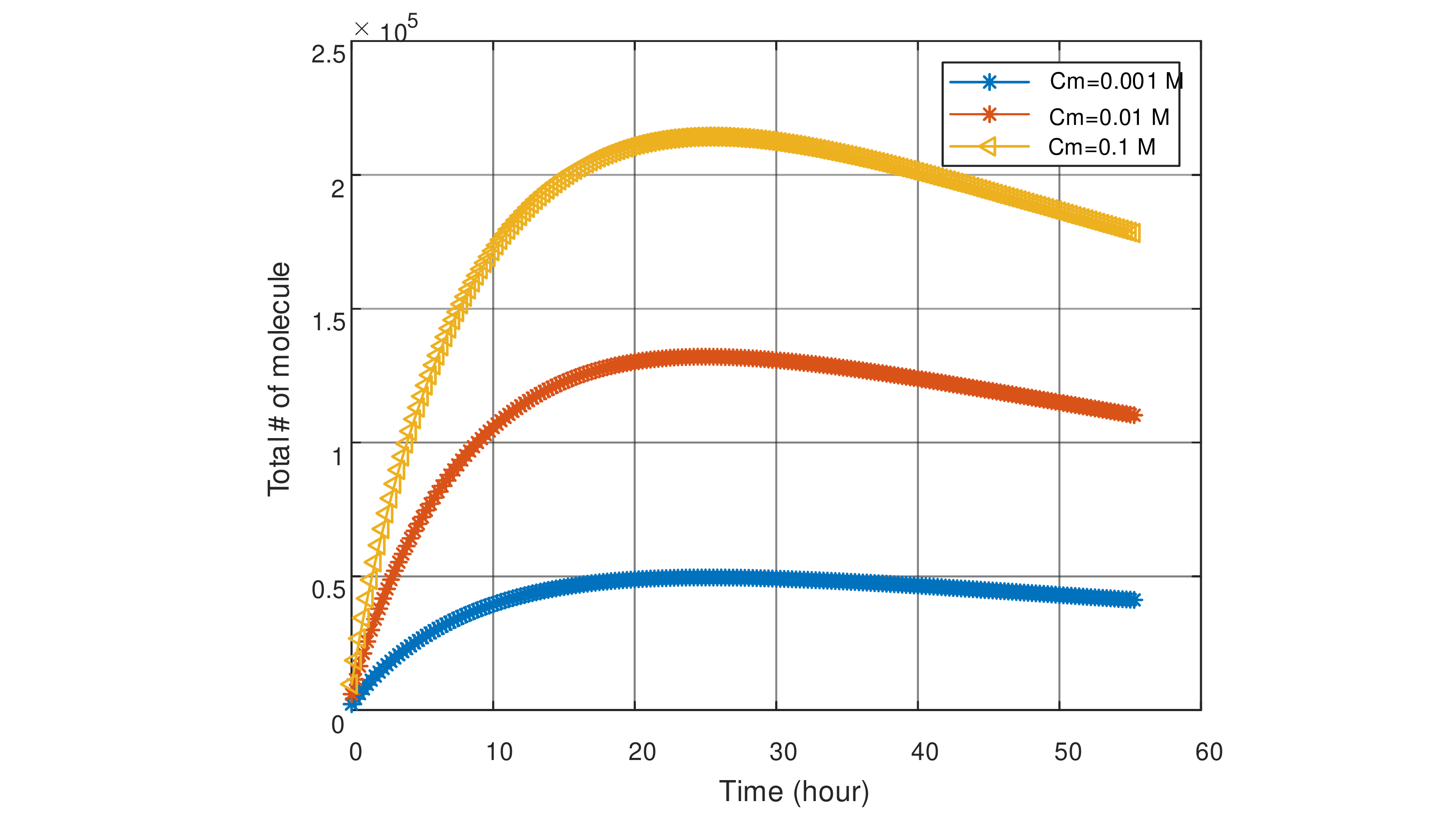}
\caption{Total number of molecules soaked into agar versus time for different IPTG droplet concentrations when the droplet makes contact with the agar at time $t$=0.}
\label{rd_mol}
\end{figure}
\begin{figure}
\centering
\includegraphics[width=3.5in]{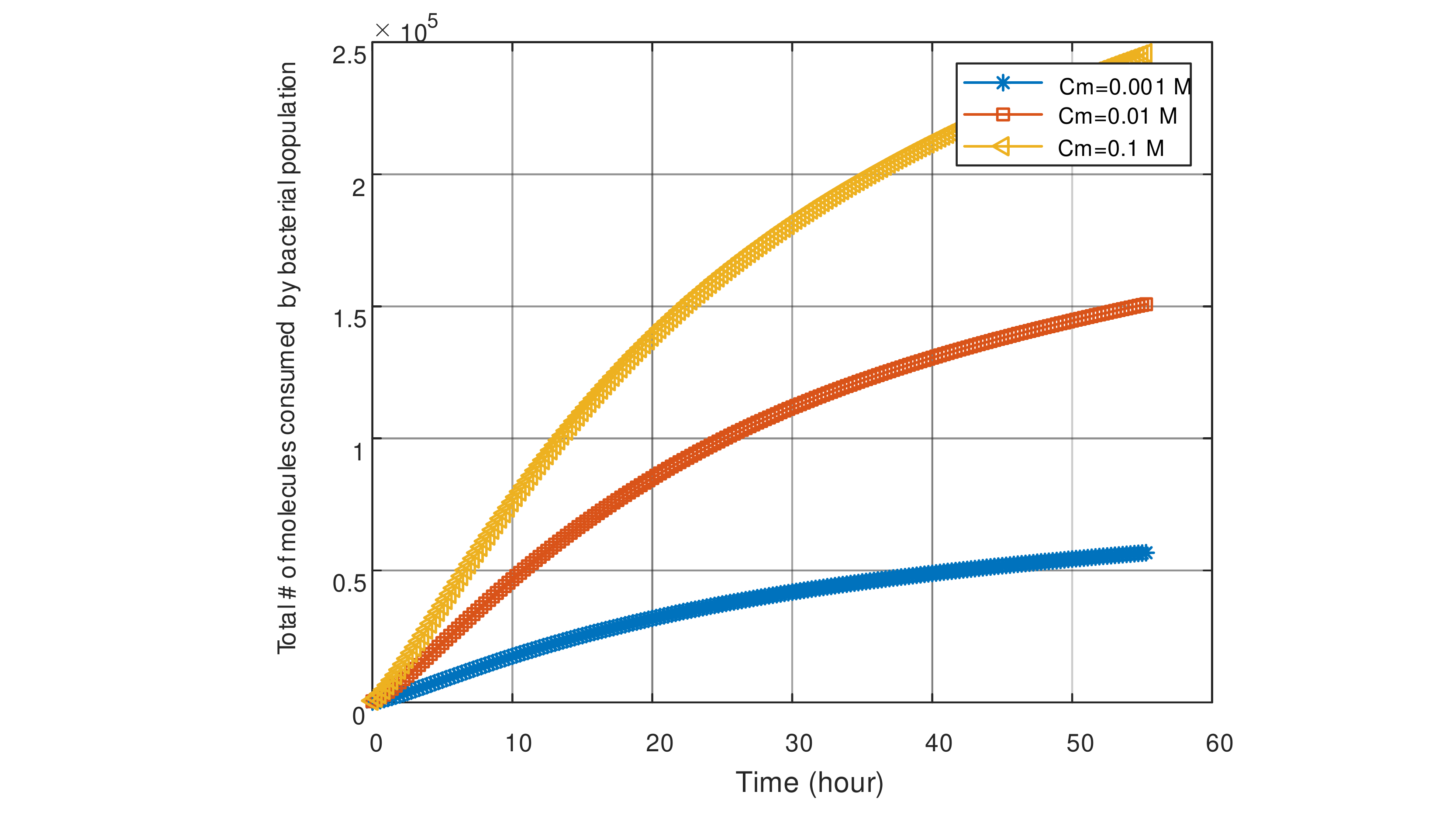}
\caption{Total number of molecules consumed by bacteria versus time for different IPTG droplet concentrations when the droplet makes contact with agar at time $t$=0 and the bacteria population is growing on the agar surface.}
\label{rd_bact}
\end{figure}
The spatio-temporal profile of the molecule concentration inside the agar (see Fig. \ref{mol_pos}) and the amount of molecule consumed by the bacteria over the surface (see Fig. \ref{bac_pos}) are plotted using the color maps. The color map provides the IPTG amount profile in 2D $(x,y)$ (integrating over z-axis for Fig. \ref{mol_pos} and \ref{bac_pos}) for 3 different IPTG droplet concentrations, $C_m=(0.001, 0.01, 0.1)$ M. Each row shows 4 time samples of the profile at \{0, 10, 25, 55\} hours from left to right. The color of each picture in Fig. \ref{mol_pos} is coded from red to blue; if the concentration is high then it is shown in red, otherwise in blue. As observed, the total amount of molecules increases first and gradually decreases as some molecules are being consumed by bacteria. The color of each picture in Fig. \ref{bac_pos} is coded from yellow to black; if the amount of molecule consumed by bacteria is high then it is shown in yellow, otherwise in black.  We observe that both the propagation radius and amount of molecules are increasing as time goes on until all molecules are absorbed by bacteria. Based on Fig. \ref{bac_pos}, we observe that the molecules do not necessarily reach the edge of the plate and the central bacteria consume more molecules. Interestingly, the radius at which the molecules arrive at the bacteria depends on the initial IPTG concentration (i.e., the amount of IPTG molecules available in the droplet at the beginning). 

The total number of  molecules within the agar versus time is shown in Fig. \ref{rd_mol} for different IPTG concentrations in the droplet. The total number of molecules increases for 20-30 hours. Then, it starts decreasing as the release ceases and/or the molecule consumption becomes dominant.  Obviously, the total number of molecules is higher for the highest IPTG concentration, as more molecules are introduced into the agar for higher concentrations of IPTG. 

The total number of molecules consumed by bacteria over the agar is depicted in Fig. \ref{rd_bact}. The total number of consumed molecules is an increasing function and reaches a level that equals the total number of molecules released into the agar.  

\section{Conclusion}

In this paper, a biophysical model of the signaling within a cell culture is proposed. The proposed model consists of three processes: droplet soaking and molecule release into agar, diffusion of signaling molecules throughout the agar, and interaction of the molecules with a growing cell population. We analyzed and characterized the molecules release rate into agar as a source term for the diffusion model. We carried out experiments using $E.\,coli$ bacteria and IPTG molecules to derive the parameters of the model and evaluate the proposed model. It was shown that the droplet area shrinks as it is soaking into the agar. We used PBS to simulate the diffusion of the particles in the agar and reaction with the bacteria over the surface and thus obtained the spatio-temporal profile of the number of molecules consumed by the bacteria. We revealed that the amount of molecules consumed by bacteria is very heterogeneous across the radius and this should be accounted for when making inferences from biological studies. In other words, the results clearly show that bacteria closer to the center can consume many more molecules. In the future, we will employ a more rigorous evaluation of the proposed models and model parameter estimation and then a biophysical model of a neuron can be designed for nano communication networks using genetically-engineered cells.

\section*{Acknowledgments}
This work was supported by the Engineering and Physical Sciences Research Council [EP/V030493/1].


\bibliographystyle{IEEEtran}
\bibliography{sample-base}


\end{document}